# Experimental and theoretical studies on visible light attenuation in water

A. Simpson$^{\$}$, A. Ludu$^{\#}$, H. J. Cho$^{\$}$ and H. Liu$^{\#}$

$^{\$}$Bethune-Cookman University, Department of Integrated Environment Science
Daytona Beach, FL 32114
and
$^{\#}$Embry-Riddle Aeronautical University, Department of Mathematics,
Daytona Beach, FL 32114

**ABSTRACT**
In this study we describe lab experiments on determining the above water reflectance $R_{rs}$ coefficient, and the water attenuation coefficient $K_d$ for fresh water. Different types of screens (totally absorbent, gray, etc.) were submerged in water (0-0.6 m) and illuminated from outside. The spectral density of the water leaving radiance was measured for different depths. The results were ran by a code which took into account the geometry of the incident irradiation, the geometry of the screen under water, and boundary conditions at the water surface provided by the radiation transfer theory. From the experimental data and our model we obtain the spectral distribution of the attenuation coefficient for fresh water and compared it with other data in literature. These experiments, performed in the Nonlinear Wave Lab at ERAU$^{\#}$ represent just a preliminary calibration of the experimental protocol. More tests with water of different degrees of turbidity, and possibly wave filed at the water surface are in progress and will be presented in a forthcoming paper.

1. **Introduction. The Radiation Theory**

Scattering and reflection of radiation in water is controlled by several mechanisms out of which the dominant ones are reflection, absorption and back scattering [1,2]. The theoretical framework of the model is supported by the radiative transfer theory for light propagation in water (dark room lab and ocean). The equation that describes the process is the radiative transfer equation which can be written for a plane geometry (that is space is divided in two homogenous half-spaces, air above and water below, by a plane horizontal interface) in the form

$$\cos\theta \frac{dL}{dz} = -c\,L + S + \int_0^{4\pi} L\,\beta\,d\Omega' \qquad (1)$$

where z is the perpendicular direction on water surface pointing downwards and $\theta$ is the polar angle between the direction of propagation of the beam and the vertical. This equation represent a spectral component of the radiative transfer balance for a given wavelength, at a given point z and at a given direction of propagation of the incident radiation.

The unknown function $L(z, \theta, \varphi, \lambda)$ is the spectral radiance defined as the amount of radiative energy traveling in the $(\theta, \varphi)$ direction in polar coordinates with respect to the z axis, per unit of area, unit of time, unit of wavelength $\lambda$, and per unit of solid angle $\Omega$. The term S represents contribution to the energy balance from the thermal radiation of the medium. The coefficient $c(z, \lambda)$ is the beam attenuation coefficient, and $\beta(z, \Omega, \Omega', \lambda)$ is the volume scattering function depending on the incident radiation direction $\Omega = (\theta, \varphi)$ and scattered directions $\Omega' = (\theta', \varphi')$ over which the integration is performed. This volume scattering function has the physical interpretation of scattered intensity per unit incident radiance per unit volume of water. Moreover, the attenuation can coefficient be broken into two parts

$$c(z, \lambda) = a(z, \lambda) + b(z, \lambda) \qquad (2)$$

where a is the absorption coefficient (responsible for loss of radiant energy by inelastic scattering and conversion into heat, and b is the scattering coefficient responsible for re-emission of radiation of different directions from the incident one. The scattering coefficient is directly related to the volume scattering function by the relation

$$b(z, \lambda) = \int \int_0^{4\pi} \beta(z, \theta', \varphi', \theta, \varphi, \lambda) \sin \theta' \sin \theta \, d\theta' d\theta d\varphi' d\varphi \qquad (3)$$

This formula represents the total scattered power per unit incident radiance and unit volume of water, and it is obtained by the integration of the scattering volume function over all directions (solid angles) generating the spectral scattering coefficient. Hals of this integral taken only over the upper hemisphere represents the backscattering coefficient, and the other hemisphere provides the forward scattering coefficient.

The spectral radiance L completely specifies the radiation field in all directions which makes it more difficult to be measured directly [2,3]. A most commonly measured radiometric quantity is the radiance E. Let us assume a certain light detector equally sensitive to photons of a given wavelength $\lambda$ traveling in any direction $(\theta, \varphi)$ within a hemisphere of directions. If the detector is located at depth $z$ and is oriented facing upward, so as collect photons traveling downward, then the detector output is a measure of the spectral downwelling scalar radiance at depth $z$, $E_d(z, \lambda)$. Such an instrument is summing radiance over all the directions in the downward hemisphere. So we can write the relation between the downwelling spectral radiance the spectral irradiation in the form

$$E_d(z, \lambda) = \int_0^{2\pi_d} L(z, \theta, \varphi, \lambda) d\Omega \qquad (4)$$

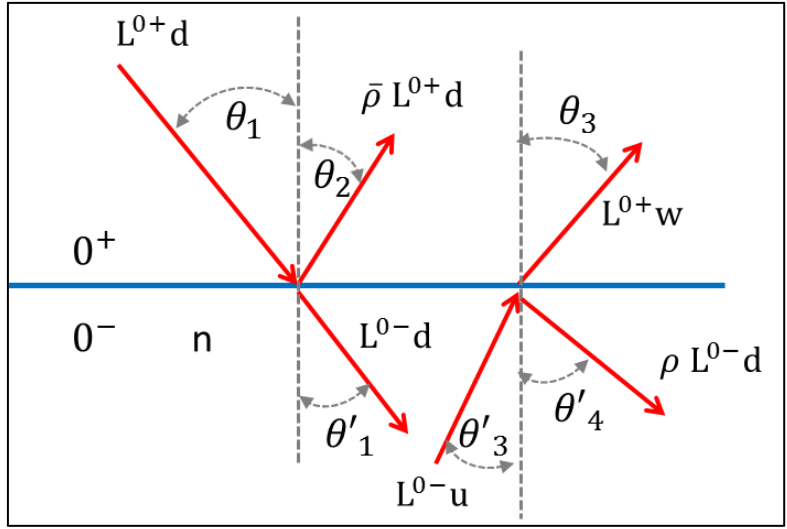

Fig. 1. The boundary conditions for radiation transfer at water surface. The index of refraction of water is n=1.34. The symbols $0^\pm$ represent: right on top, and right below the water surface, respectively. The radiance is represented by $L_{u,d}^{0\pm}$ function of the point where it is evaluated and the direction of the light beam (subscript u stands for upwelling, and d for downwelling radiance). $\rho$ and $\bar{\rho}$ represent the Fresnel reflectivity coefficients, water-to-water (internal Fresnel reflectance), and air-to-air (or air-water Fresnel reflection at the interface), respectively.

If the same instrument is oriented facing downward, so as to detect photons traveling upward, then the quantity measured is the spectral upwelling scalar radiance. The spectral scalar radiance is the sum of the downwelling and upwelling components [4,5]

$$E(z,\lambda) = E_d(z,\lambda) + E_u(z,\lambda) = \int_0^{4\pi} L(z,\theta,\varphi,\lambda)d\Omega \qquad (5)$$

These equations provide the boundary conditions when we try to solve the integro-different ail Eq. (1) across the water surface. In Fig. 1 we present the main notations with respect to radiances, irradiances and $\rho$ and $\bar{\rho}$ representing the Fresnel reflectivity coefficients, water-to-water (internal Fresnel reflectance), and air-to-air (or air-water Fresnel reflection at the interface), respectively. The angles are evaluated with respect to the local vertical and prime angles are in water and follow the Snell law of refraction. Radiation energy balance calculated right above the water surface is denoted with superscript $0^+$ and below the surface with $0^-$.

In order to solve Eq. (1) we need to make the substitutions $\mu = \cos\theta$ and

$$\tau(z) = \int_0^z c(z')dz'$$

The new independent variable $\tau$ is called optical depth. With this substitution we can search solutions L for Eq.(1) by decomposing the radiance L in the upwelling and downwelling parts, as we shown in Fig. 1. With these notations, the formal solutions of Eq. (1) can be written [2,4,5]

$$L_u(\tau,\mu,\lambda) = L_u(0,\mu,\lambda)e^{-\frac{\tau}{\mu}} + \frac{1}{\mu}\int_0^\tau e^{-\frac{\tau-\tau'}{\mu}}\left(S(\tau',\mu,\lambda) + \int_0^1 L_u(\tau',\mu,\lambda)\beta(\tau,\mu,\mu',\lambda)d\mu'\right)d\tau' \qquad (6)$$

for $\mu > 0$ and a similar integral for the downwelling radiation and $\mu < 0$.
At the water interface, the boundary conditions can be written as
$$L_u(0,\mu,\lambda) = S(\mu,\lambda) + |\rho(0^-,\mu,\lambda)|^2 L_d(0,\mu,\lambda) \qquad (7)$$
where S is the thermal emissivity of the water surface, and R is the reflectance of the light from water going upwards, expressible in terms of the Fresnel power coefficients. A similar relation can be written above the water surface
$$L_d(0,\mu,\lambda) = S(\mu,\lambda) + |\bar{\rho}(0^+,\mu,\lambda)|^2 L_u(0,\mu,\lambda) \qquad (8)$$
In the following in this article we neglect the source term, $S(\mu,\lambda) \approx 0$, because in the visible spectrum range and regular ocean temperature the thermal emission of water is negligible, unless we would study layers of phosphorescent plants or animals in night time.

## 2. The Experimental Model

There are several simple models in literature, all based on exponential attenuation of light [1,2,3]. The main assumption made in these simplified models is that the light falls on the water surface perpendicular and reflects back the same way, so the results do not depend on the geometry of the experiment, or on the reflection/refraction angles. In our model we take into account the geometry because the sources are placed at finite distance from the screen, and not like the sun at infinite distance. A circular screen of radius R is placed at a depth h parallel to the water surface. The tank is rectangular with the width L and the sensor is placed at the center at a height d above water surface, pointing vertically downwards towards the center O, Fig. 2. In the experiment we placed four identical wide sources of light around the vertical symmetry axis, and we consider in the model that this configuration can be well approximated with a circle of such light uniformly placed around the axis.

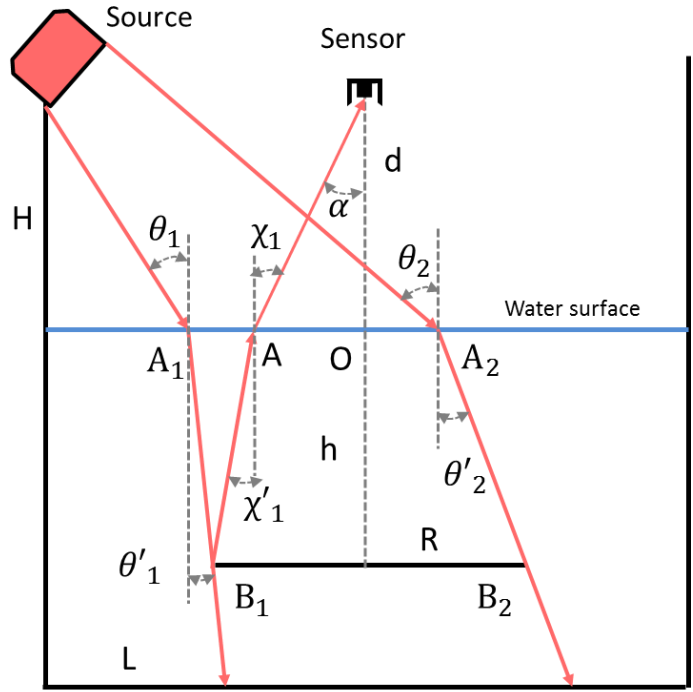

Fig. 2. The energy flux from a source placed at height H is received by the screen only through an area defined by the angles $\theta_1$ and $\theta_2$. The reflected radiation from a screen placed at depth h leaves the water under angle $\chi_1$ limited by the maximal angle of view $\alpha$ of the sensor.

First, we note that since the sensor measuring the reflected light is placed at d=0.5 m above the water surface and points vertically downwards having a total field angle of view of α=27° the sensor detection radius under water is given by

$$r_{measured} = \left(2R - h + \frac{h}{1.33}\right) \cot 27^0$$

where n=1.33 is the refraction index of water. Working with a column of $h = 0.66$m, and with circular screens of radius R=0.25 m (from point $B_1$ to $B_2$ in Fig. 2) the sensor would catch more field than the total screen surface only if placed no deeper than $h_{critical}$=0.575m. This criticality does not occur in our experiments so we can consider that the sensor beam detects always only the surface of the screen underwater, and nothing coming directly from the bottom or walls (of course neglecting parasite reflections).

We assume that the sources are circularly distributed above the tank, and around the vertical symmetry axis dropping from the sensor. In Fig. 2 where we sketch the experimental setup we show just the light at the left, but we consider a uniformly distribution of such sources around the axis, so imagine a similar red lamp also drawn on the right upper edge of the tank in Fig. 2. From an elementary source placed at height H above water the screen receives only a limited solid angle of radiance. We have to take into account only the light field between the points $A_1$ and $A_2$ from Fig. 2. The positions of these points and their angles can be calculated function of the depth h of the screen, the radius of the circular screen R, the width of the water tank L, and the height of the source above the water H. The resulting expression for both limits reads

$$\frac{\sin \theta_{1,2}}{\sqrt{n^2 - (\sin \theta_{1,2})^2}} = \frac{L}{2h} \mp \frac{R}{h} - \frac{\sin \theta_{1,2}}{\sqrt{1 - (\sin \theta_{1,2})^2}} \qquad (9)$$

By solving numerically these equations we obtain the limitation of the incoming light solid angle downwelling the water surface. We found, for example:

$$\theta_1(h) = \frac{0.676 + \sqrt{0.633 + 2.275\, h}}{10\, h} \qquad (10)$$

and a similar expression for $\theta_2(h)$. Let $E_0$ be the radiance at the water surface received from these light sources. We consider as initial condition on top of the water the downwelling irradiation $L_d(0^+, \mu, \lambda)$ with $\cos^{-1} \theta_1 \leq \mu \leq \cos^{-1} \theta_2$ and $450 \leq \lambda \leq 950\ nm$.

At the water surface the light beam radiance $L_d(0^+, \mu, \lambda)$ is split in two parts: the reflected beam (radiance leaving the water surface) and the transmitted beam. A part of the water surface reflected beam arrives to the sensor. The rest of the reflected beam is scattered downwards in all directions and it is lost for the detector. We denote this scattered, and not measured directly radiance by $L_1$. The calm water surface reflection is rather specular, but we avoid the direct reflection direction from lights to fall on the sensor. The sensor receives only the Lambertian part of water surface reflection, and also parasite light coming from several other surfaces reflecting light in an averaged way, independent of the distance water surface-sensor.

The transmitted beam radiance $L_d(0^-, \mu', \lambda)$ travels inside the water column, and interacts with water molecules, solutes and other particles. This interaction has two contributions to the transmitted beam: the beam is continually attenuated (exponentially by the Beer-Lambert law [1-11]), and a part of the beam is scattered around. The attenuated beam reflects finally on the submerged screen and travels upwards. This upwelling irradiation is again attenuated and scattered on its way to the water surface. When touches the water surface from below it losses energy by a water-to-water reflection, and when emerges in air also through the Starubel-Fresnel $n^2$ law of radiance [13]

$$\frac{L_w^{0^+}}{L_u^{0^-}} = (1 - \rho)\frac{1}{n^2} \qquad (11)$$

where we use the notations from Fig. 1, namely the left hand side of the equality is the ratio between the water leaving irradiation above water level, $L_w^{0^+}$, over the upwelling irradiation coming through water from the screen.

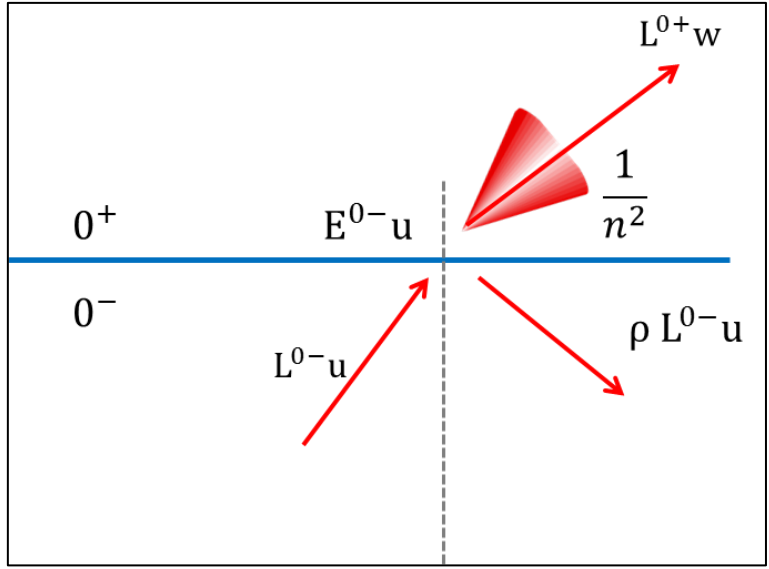

Fig. Xa.

Here again $\rho$ (sometimes denoted $\rho_F$) is the coefficient of reflectivity from water into water, or the internal Fresnel reflectance, and n=1.34 is water index of refraction. The ratio in Eq. (11) describes how energy balance works for the radiance when it travels upwards the water surface.

In the following we will construct the most important optic property of the model, the above water reflectance

$$R_{rs} = \frac{\pi\, L_w^{0+}}{E_d^{0+}} \qquad (12)$$

where $L_w^{0+}$ is the water leaving radiance measured above the water surface (namely what radiation coming from water would be measured by a detector pointing downwards), and $E_d^{0+}$ is the downwelling irradiance above the water (namely the total upwelling radiation coming from a perfect reflective screen placed above water by the sensor pointing, again, downwards). This coefficient is sometimes denoted $\rho_w$. The formula is explicitly represented in Fig. Xb.

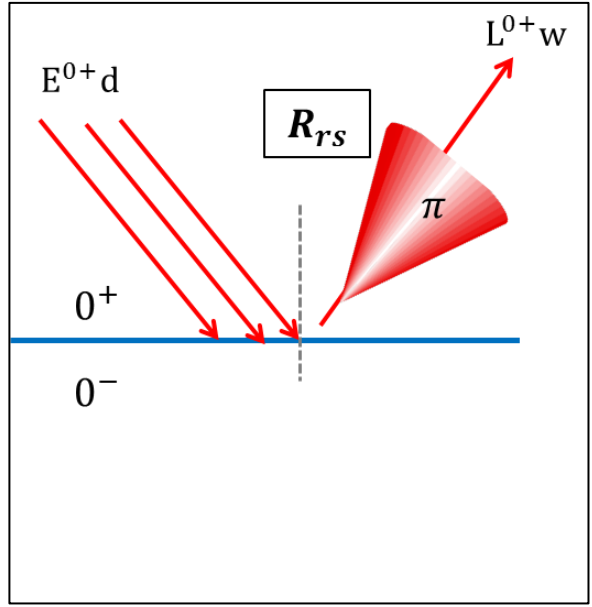

Fig. Xb. Energy balance above the water surface: The total energy flux input is the downwelling irradiance flux $E_d^{0+}$. The output of the $R_{rs}$ formula is the total radiance leaving the water, $L_w^{0+}$, multiplied by $\pi$ in order to integrate the emerging energy all over the upper hemisphere above water.

The next energy balance we are interested is what happens right below the water, see Fig. Xc. This balance is described by the irradiance reflectivity just beneath the surface (or clear water irradiance reflectivity) defined as

$$R^{0-} = \frac{E_u^{0-}}{E_d^{0-}} \qquad (13a)$$

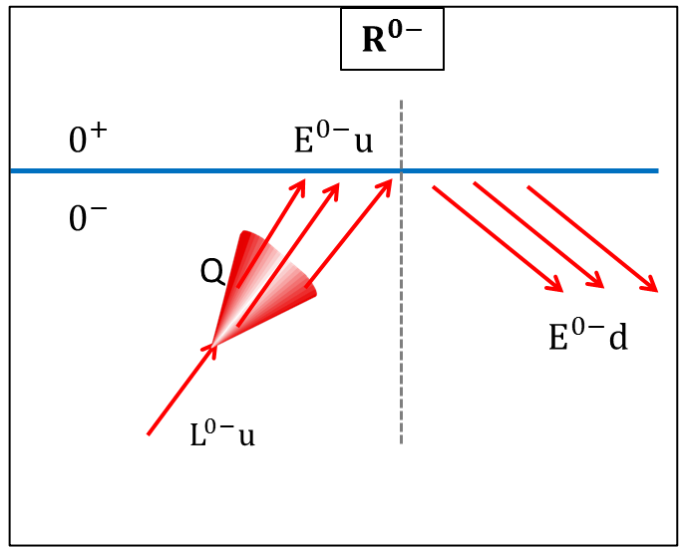

Fig. Xc. Energy balance underneath the water surface: The total input is the upwelling radiance $L_u^{0-}$ multiplied with a factor Q equal to a certain arbitrary solid angle. The resulting quantity is, by definition

from Eq. (4) the upwelling irradiance flus under water $E_u^{0-}$. The output of the $R^{0-}$ formula is the total irradiance going back in water, $E_d^{0-}$.

This formula describes what amount of the upwelling irradiance under water is reflected by the water-to-air surface, back into the water. The reflectivity just beneath the water has a strong dependence on the angles, so there is the tradition to describe it rather in terms of the upwelling radiance. We define a geometric factor $Q(\theta, \varphi) = E_u^{0-}/L_u^{0-}(\theta, \varphi)$ with dimensions of solid angle (sr). With this transition we can re-express the irradiance reflectivity just beneath the surface in the form

$$R^{0-} = Q \frac{L_u^{0-}}{E_d^{0-}} \qquad (13b)$$

And in this dimensionless form is also called inherent ocean surface radiance to the irradiance reflectance, see Fig. Xc.

The final relation we need in order to build a rigorous measurement procedure is the detailed energy balance when radiation crosses the water surface, in both directions. We elaborate partially on this topic in Eq. (11) and in Figs. 1 and Xa., but now we will take into account both upwelling and downwelling radiation. The balance is graphically explained in Fig. Xd.

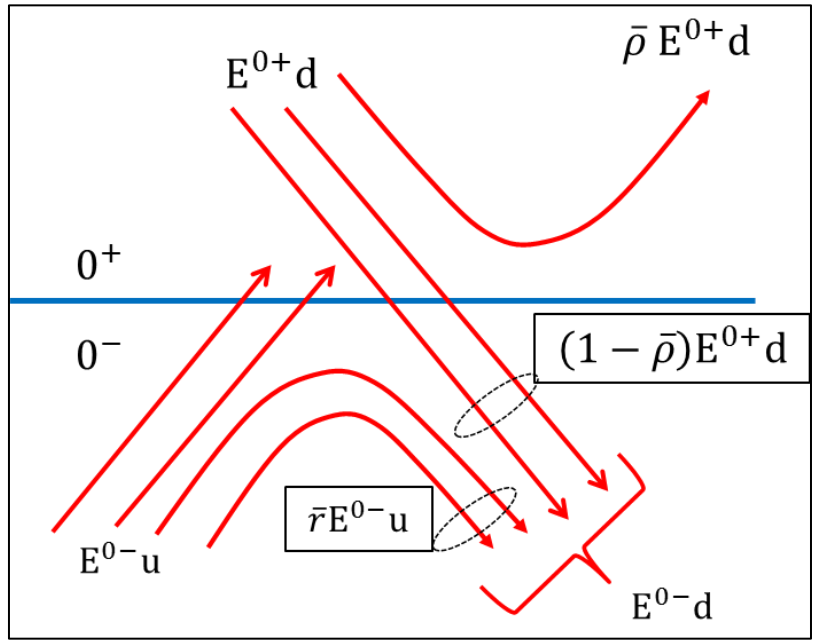

Fig. Xd.

The downwelling irradiance $E_d^{0+}$ above the water is split at crossing the surface into a part reflected back to the sensor, $\bar{\rho} E_d^{0-}$ and a part which transverses the water. Here $\bar{\rho}$ is the air-water Fresnel reflection at the interface. At the same time the upwelling radiation below the water $E_u^{0-}$ is also split into a reflected component which returns to the water column, $\bar{r} E_u^{0-}$, and another component crossing the surface upwards to the detector. The water-air reflection coefficient $\bar{r}$ describes the reflectivity beneath water of this part of the radiation and has an average value of 0.48. The total downwelling irradiance beneath water is now the sum of these two components, see Fig. Xd, namely

$$E_d^{0-} = E_d^{0+}(1 - \bar{\rho}) + \bar{r}\, E_u^{0-}$$

By using Eq. (13a) we have

$$E_d^{0-} = E_d^{0+}(1 - \bar{\rho}) + \bar{r}\, R^{0-} E_d^{0-}$$

which equation can be re-written in the form

$$E_d^{0-}(1 - \bar{r}\, R^{0-}) = E_d^{0+}(1 - \bar{\rho})$$

or simply as a ratio

$$\frac{E_d^{0+}}{E_d^{0-}} = \frac{1 - \bar{r}\, R^{0-}}{1 - \bar{\rho}} \qquad (14)$$

In the following we combine all energy balance relations we constructed Eqs. (11-13) in the form

$$R^{0-} = \frac{E_u^{0-}}{E_d^{0-}} = \frac{Q}{\pi} \frac{n^2(1 - \bar{r}\, R^{0-})}{(1 - \bar{\rho})(1 - \rho)} R_{rs}$$

This is the most important optical relation for ocean radiation balance because it describes the relationship between the above water reflectance $R_{rs}$ (which is measured with detectors from the sky) and the irradiance reflectivity just beneath the surface $R^{0-}$ (which cannot be easily measurable directly, but contains the main physics of the process in terms of inherent optical parameters of the ocean [1,4,8,12,14].

3. **Qualitative analysis of data and discussions on relevance of the model**

The radiance leaving the water surface by reflection air-to-air above the surface is:

$$L_R = L_u(0^+, \mu, \lambda) = \begin{cases} L_d(0^+, -\mu, \lambda) R_{\text{surface}}(\mu, \lambda) & \text{if } 0 < h \text{ (screen submerged in water)} \\ 0 & \text{if } h = 0 \text{ (screen above water)} \end{cases} \qquad ()$$

were h is the depth of the screen. The water reflectivity coefficient $R_{\text{surface}}$ includes the contribution of multi path signals like the background reflections, residual light, and it depends on the wavelength and the angle ($\theta = \cos^{-1}\mu$) of the corresponding incident beam.

The energy balance above water is given by

$$L_d(0^+, \mu, \lambda) = L_d(0^-, \mu', \lambda) + L_u(0^+, \mu, \lambda) + L_1 \qquad ()$$

By combining all these contributions, the irradiation arriving at the detector at an angle $\chi_1 \leq \alpha$, where $\alpha$ is the vision (half-) angle of the sensor is given by the final formula

$$L_{screen} = L_t e^{-K_d h} R_{screen} R_0(h) e^{-K_d h}$$

where $K_d > 0$, $0 \leq R_{screen} \leq 1$, $0 \leq R_0(h) \leq 1$, are the water attenuation coefficient (wavelength dependent), screen reflection coefficient, and the function $R_2(h)$ which takes into account the multi path signals generated by secondary reflections, absorption of light, thermal effects, etc.

The sensor also receives some of the scattered light from the transmitted beam through the water column. Considering an infinitesimal layer of water of height dz, placed at a depth $0 \leq z \leq h$, a fraction $dL = L_t e^{-K_d z} R_{bs} dz$ of light is backscattered, depending on a mean backscattering coefficient $R_{bs}$. The total amount of light backscattered during the beam's trip to the screen will is

$$L_{bs} = L_t R_{bs} \int_0^h e^{-2K_d z} dz = \frac{L_t R_{bs}(1 - e^{-2K_d h})}{2K_d}$$

The total amount of light received by the sensor is the sum of these four contributions $L_{sensor} = L_R + L_1 + L_{screen} + L_{bs}$, namely

$$L_{sensor} = L_1 + L_0 \left[ R_{surface} + R_t\, R_{screen}\; e^{-2K_d h} + \frac{R_t R_{bs}(1 - e^{-2K_d h})}{2K_d} \right]$$

or simply

$$I_{sensor} = I_1 + I_0 \left[ R_{surface} + \frac{R_t R_{bs}}{K_d} + e^{-2K_d h} R_t\; R_{screen} - e^{-2K_d h} R_t \frac{R_{bs}}{K_d} \right] \quad ()$$

The values for $L_0$ and $L_1$ can be measured directly. $L_0$ is evaluated by combining and re-scaling the light measured upwards vertically and the light $L_{exp5}$ measured by pointing the sensor towards one light source (at $30^0$ from vertical) and then multiplied by four (the number of light sources), all in all we denote it:

$$L_0 = 4\, L_{exp5}$$

The parameter $L_1$ is obtained by drying the tank and measuring the light $L_{exp4}$ reflected by the walls and the rest of background, without screen or water, that is

$$L_1 = L_{exp4}$$

Next preliminary experiments measure the light reflected by the empty water column (no screen) $L_{exp3}$ and the light reflected by the screen only (screen placed on top of water surface), $L_{exp1}$. From here it is easy the determine two more coefficients

$$R_{screen} = \frac{L_{exp1} - L_1}{L_0}$$

and obtain the relation

$$R_{surface} = \frac{L_{exp3} - L_{exp4}}{L_0} + R_t \frac{R_{bs}\left(-1 + e^{-2K_d h_{max}}\right)}{2K_d}$$

which, in the experimental conditions we used ($h_{max} = 0.66$m, $R_{bs} \ll R_{fs}$ because of clear water) can be approximated very well with

$$R_{surface} \approx \frac{L_{exp3} - L_{exp4}}{4 \, L_{exp5}}$$

We present below the results of these preliminary light intensity measurements versus wavelength:

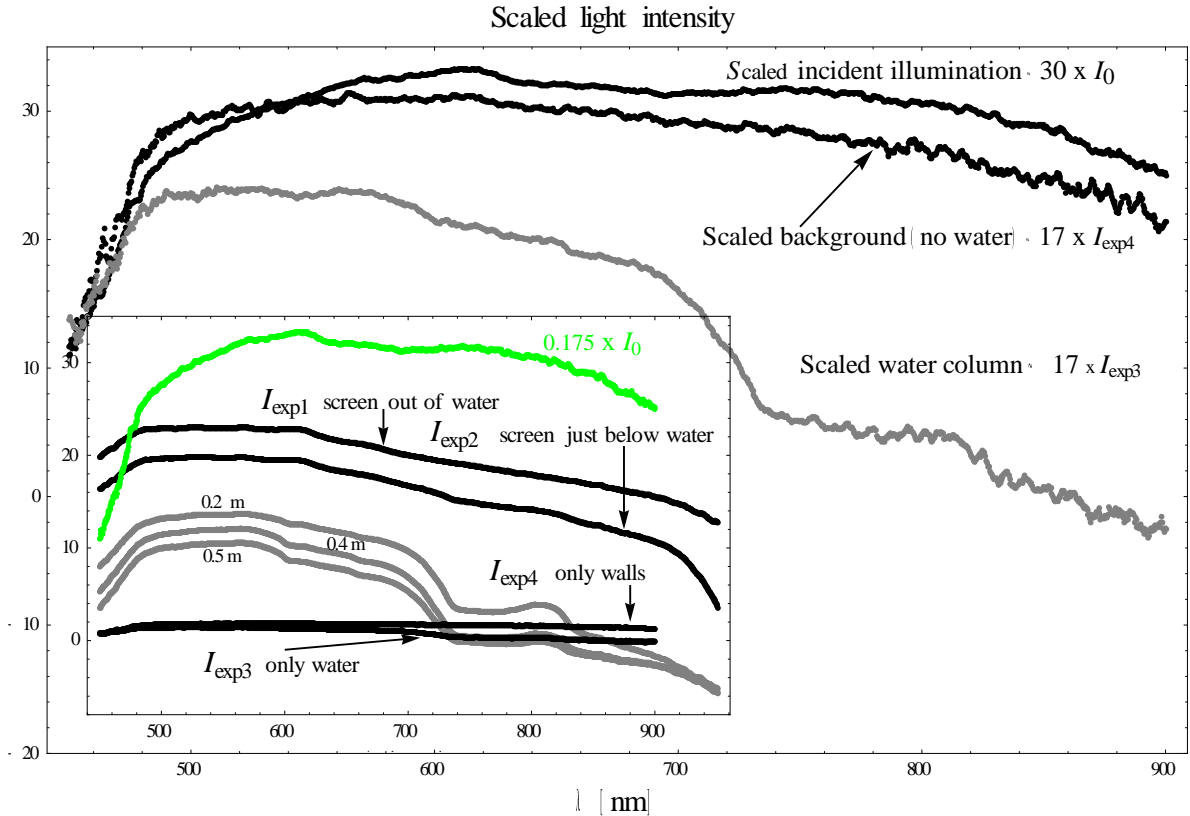

**Fig. 3 INSET:** <u>Solid black curves:</u> $L_{exp\,1}$: screen placed on top of the water surface (screen reflectivity), $L_{exp\,2}$: screen placed just below the water surface (water plus screen reflectivity, but little attenuation), $L_{exp\,3}$: only the water column and no screen, and $L_{exp\,4}$ parasite reflection from tank walls and surroundings. <u>Green curve:</u> represents the normalized incoming light from one source, scaled by a factor of 0.175. <u>Gray curves:</u> screen placed at three different depths. **LARGE WINDOW**: re-scaled

incident illumination ($L_{0/normalized}$, the highest curve), re-scaled background reflection from walls ($L_{exp3}$ middle curve), and re-scaled reflection from water surface ($L_{exp3}$ gray, lowest curve) all showing same type of behavior if rescaled to the same values. Water has a drop in reflectivity toward large wavelengths.

From Fig.3 we note that the incident spectrum can be considered uniform distribute over the interest wavelengths. Anyway, we take into account the measured value of $L_0$ for every wavelength and for every evaluation of the coefficients. In the main window in Fig. 3 we plot that the parasite light reflected by the background and walls ($L_{exp4}$) rescaled at 1:250. We note that its spectrum is very similar to the total incident light ($L_0$) which makes sense since the walls are covered in dark material with constant absorption spectrum, so they should reflect back the same spectrum.

From the reflection of light from the screen above water ($L_{exp1}$), and reflection from the submerged screen right under water surface ($L_{exp\,2}$) we notice that

$$L_{exp\,2} \approx L_{exp\,3} + L_{exp\,1},$$

up to an additive re-scaling of the sensor data showing a good agreement with the model. We also note that $L_{exp\,3} \leq L_{exp\,4}$, showing that the reflectiveness of water surface is in the same range with background parasite reflections (tank walls).

In order to compare this value with the theory we use electromagnetic radiation calculations where the reflectivity coefficient for natural day light shining straight down on water (our simplified model) is given by

$$R_{\text{surface theory}} = \left(\frac{n-1}{n+2}\right)^2 \sim 0.00982,$$

where n is the water index of refraction which is in the range of 1.33, and we compare this value with the experimental ones [2,3].

### 4. Experimental result and interpolation with theory

There were two series of experiments performed with a black screen, and a gray screen. The reflected light signal measured by the sensor was recorded and studied versus depth of the screen (h between 0 and 0.58m) and wavelength from 450 to 950 nm. The water used was clear tap water. The results are presented in Figs. 4a (black screen) and Fig. 4b (gray screen), respectively.

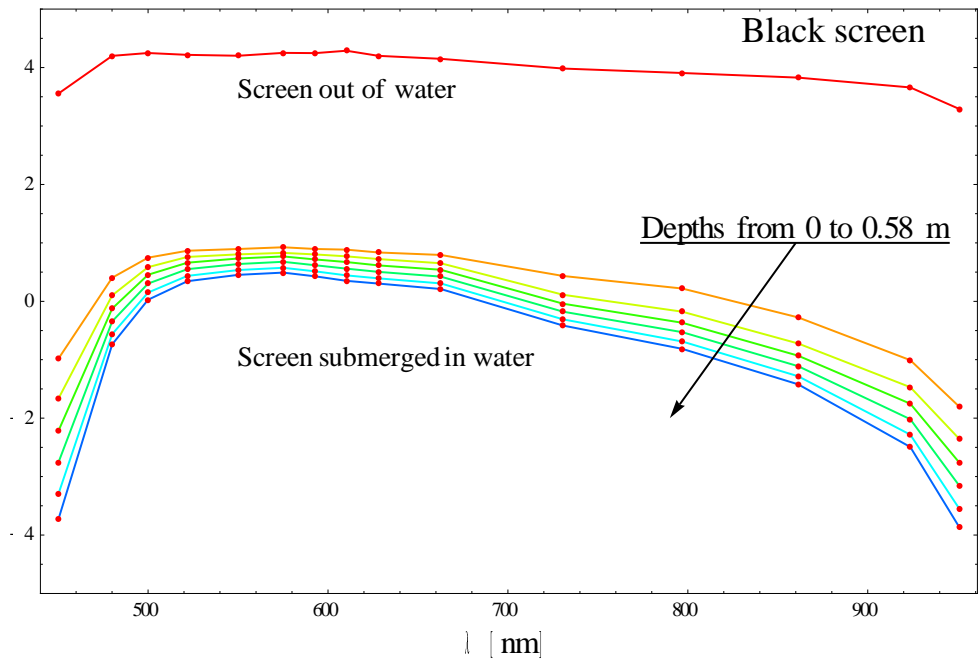

***Fig. 4a.*** *Experimental results for light intensity (relative units) with black screen*

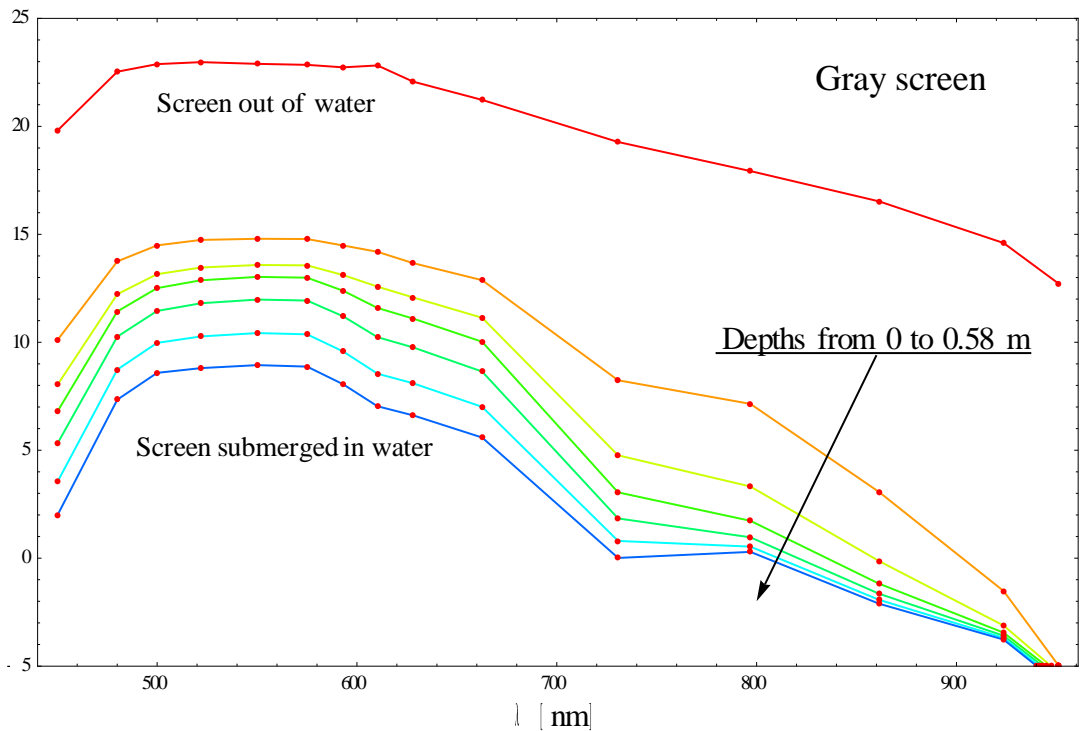

***Fig. 4b.*** *Experimental results for light intensity (relative units) with gray screen*

The experimental results were compared with the theoretical model by interpolating Eq. (9) with the experimental data and using a nonlinear regression algorithm of minimization with four free parameters $K_d$, $R_{screen}$, $R_{surface}$, and $R_t$ for each wavelength and for each depth. The results are shown in Figs. 3. There were 7 different depths and 1490 wavelength points considered. Moreover, after each interpolation between experiment and theory, the values obtained from the parameters $R_{screen}$, $R_{surface}$, and $R_t$ were compared and adjusted (averaged) with their values obtained in the preliminary experiments, and presented in Eqs. (2-5). The coefficients $I_1$ and $I_0$ were given by the preliminary experiments and the equations in the beginning of section 3. At this stage we neglected the value of the backscattering coefficient (water was very clear), $R_{bs} \approx 0$ and also neglect at this stage the other nonlinear effects by choosing $R_0(h) \approx 0$.

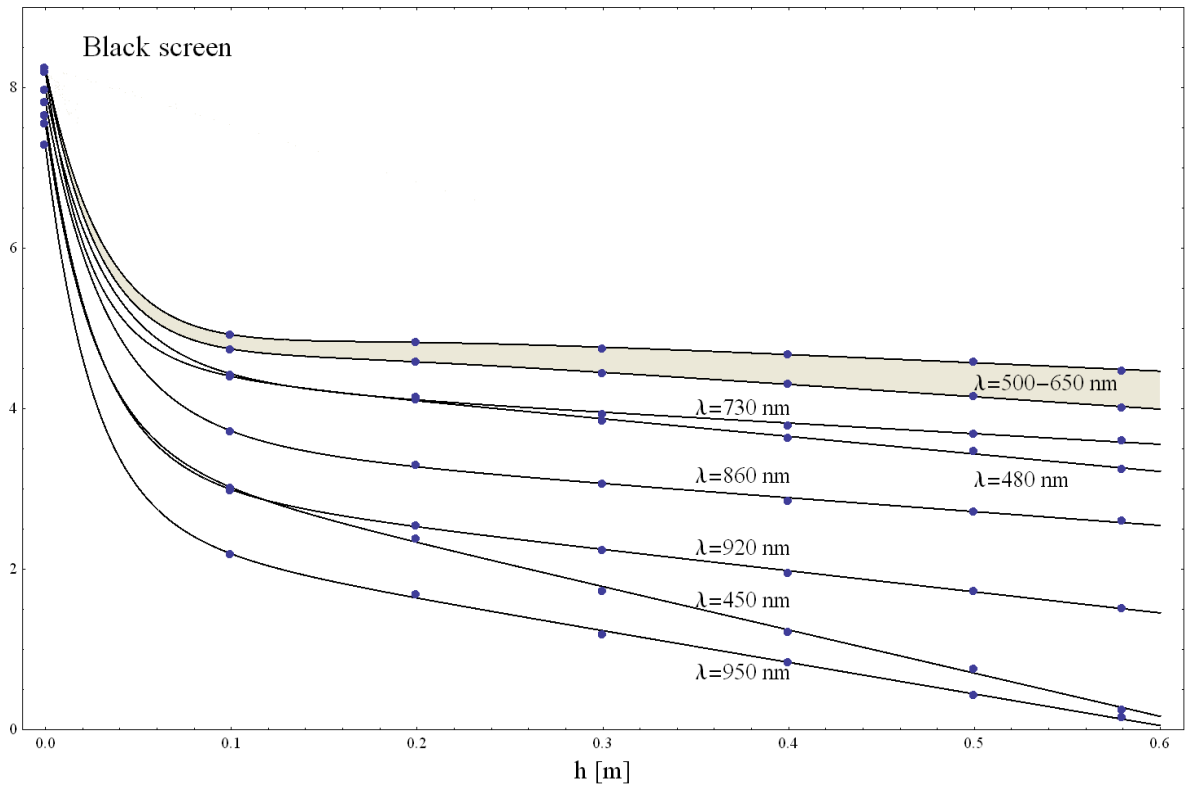

***Fig. 5a.*** *Intensity of light (in normalized relative units) reflected by the black screen submerged in water, for different depth and different wavelength. Solid lines represent theory and dots represent some key experimental points (not all 1490). The gray shaded suggests all attenuation curves laying in the middle visible spectrum (green to orange), where the attenuation is weakly dependent on wavelength and curves are very similar. The IR and UV limits are the strongest attenuated.*

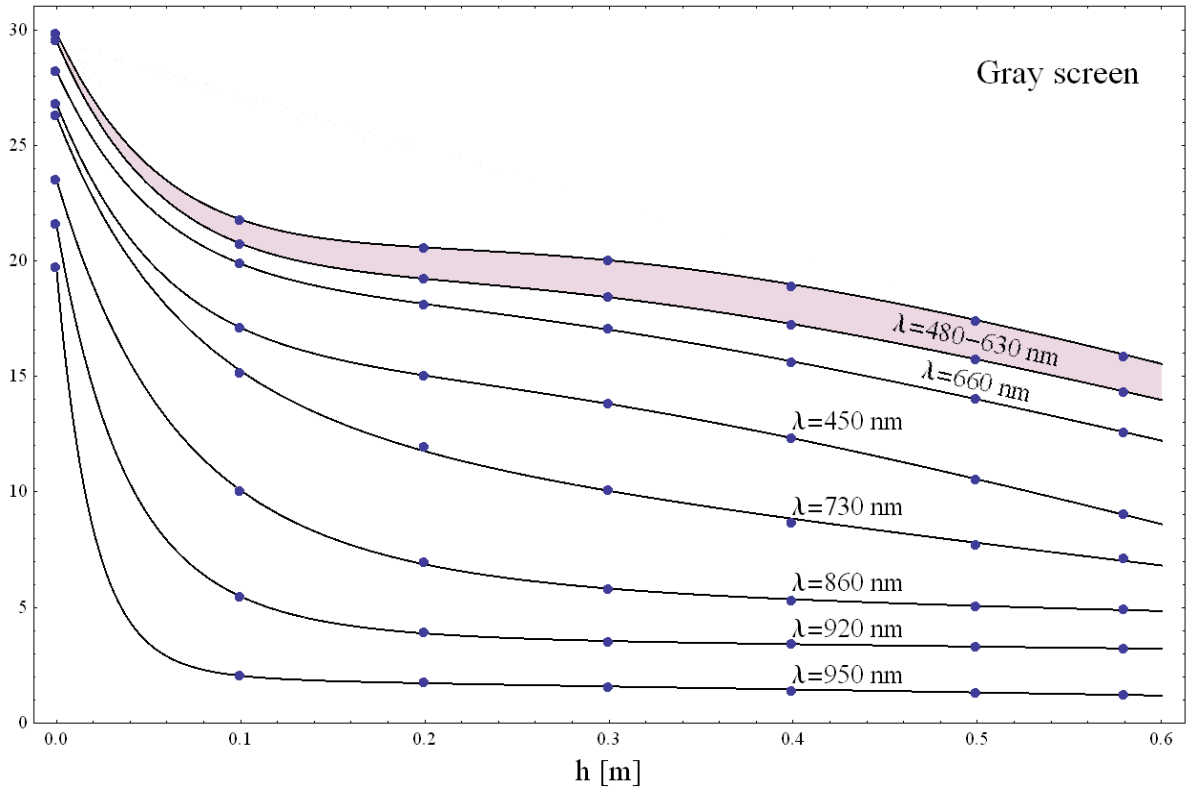

***Fig. 5b.*** *Light reflected by the gray screen submerged in water, for different depth and different wavelength. Same notations and legend as in Fig. 5a. As opposed to the black screen, here only large wavelengths have strong attenuation.*

### 5. **Analysis of results**

The theoretical model provides a good fit with experimental results, see Figs. 5 and 7. Once we obtained the best fit for Eq. (9) we can extract the coefficients. The results are presented in Fig. 6a for the black and 6b for the gray screen. The attenuation coefficient $K_d$ values from the black screen measurements have some peculiar features which we discuss in the followings. Because the black screen reflects little amount of light to the sensor, a large contribution of what the sensor records in this case comes from the water column, and of course the background.

From Fig. 6a we note that for wavelengths larger than 670 nm the $K_d$ values have a good agreement with literature [3,4]. Shorter than 670 nm wavelengths show some oscillations in $K_d$ which may not be related to attenuation though. The resulting values for the total background illumination

$$L_1 + L\left(R_{surface} + \frac{R_t R_{bs}}{K_d}\right)$$

consists in background light, Lambertian water surface reflection, and a background coming from water column backscattered light (the $I_1$ term also contain in it the calibration of the sensor) show an almost iso-spectral behavior whose constancy brings confidence to the model. Its two negative tails at very low and very high wavelengths are solely related to sensor output calibration.

From the present model is impossible to obtain independent values for the screen reflectance and transmission coefficients. We can only infer the values of the product $R_t R_{screen}$. From Fig. 6a we note that the water column transmits better light at lower wavelengths, where also the screen is more reflective, and this is in concordance with lower attenuation for these wavelengths. The transmission and reflectance have opposite minima and maxima with respect to the attenuation coefficient, which makes also sense. Also, since the screen is black and has a low reflection coefficient it seems that the forward scattering coefficient may have some non-negligible contribution in this term. Finally, the $R_t \frac{R_{bs}}{K_d}$ term shows negative and positive behavior which is explained by the balance between back and front scattering of light: short wavelength light is rather back scattered, and the opposite for large wavelengths. At short wavelengths water transmission and screen reflection dominate the picture, while at long wavelengths water column forward scattering dominates. In order to double check this behavior we expanded the solution in Taylor series in h around h=0, and analyze the spectral dependence of the first three series coefficients. The lower order powers of h (responsible for small depth behavior) are practically controlled by the screen reflection, while the higher powers of h (responsible for the deep depth behavior) show more dependence on the forward scattering process.

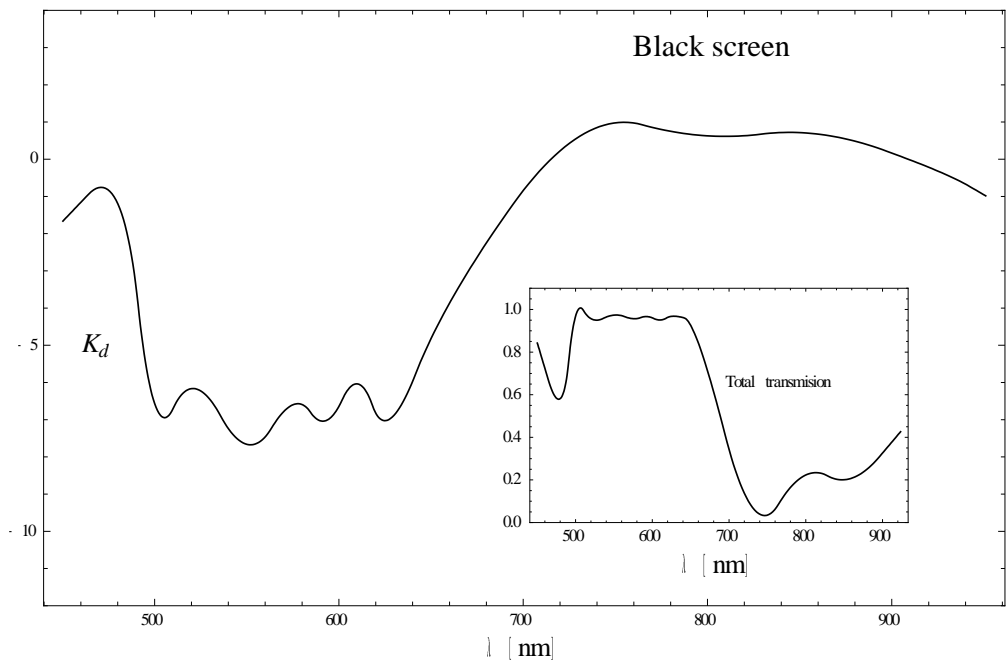

*Fig. 6a. Results of the interpolation of model on experimental results from the dark screen: attenuation coefficient ($K_d$ solid line), the transmission coefficient multiplied to difference between screen reflectivity and forward scattering coefficient, dashed), the transmission coefficient multiplied to difference between the forward scattering coefficient and the backward scattering coefficients, dotted-dashed), and background illumination ($I_1$, dotted).*

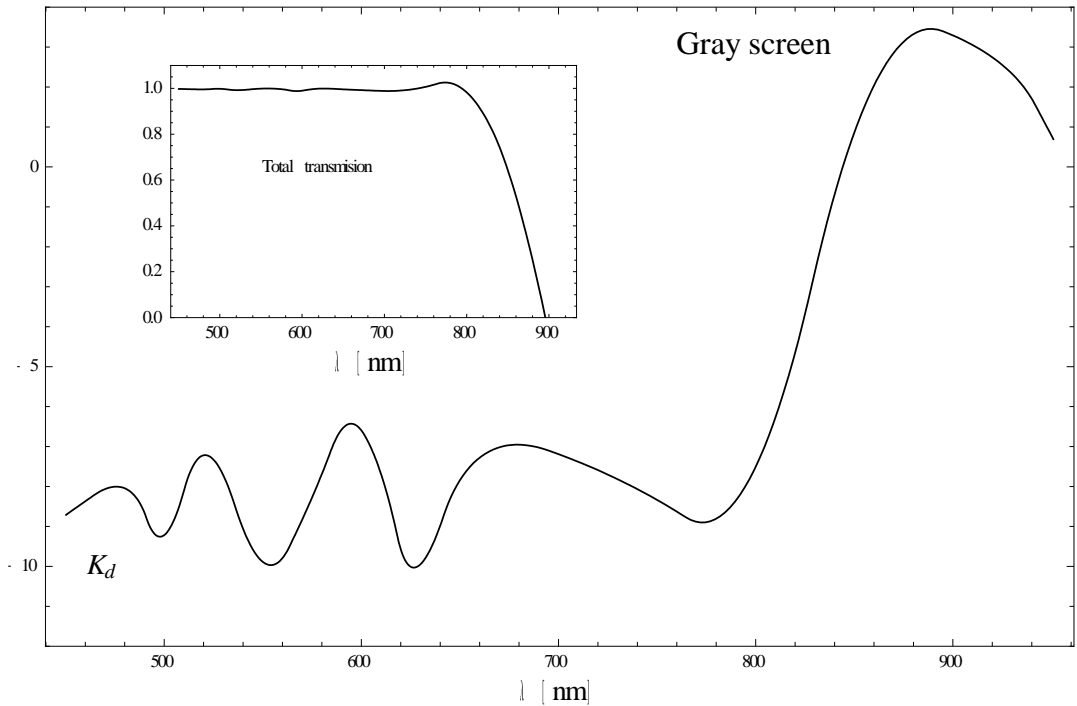

*Fig. 6b*. *Results of the interpolation of model on experimental results from the gray screen. The symbols have the same meaning as in Fig. 6a.*

In Fig. 5b we present the results for the gray screen.

The attenuation coefficient $K_d$ extracted from these experiments is presented in Figs. 6,7. Its values are close to the values obtained literature, [5-11], and the general behavior of different terms is the same as in the black screen case. In the short wavelength range the dominant process is reflection on the screen and back scattering in water. One can note this from the large negative values of the scattering term. At longer wavelength this term becomes positive showing the reverse in the dominance.

An observation is that the contribution of the scattering in the water column is rather weakly dependent of the screen position for short wavelengths. At long wavelengths, the scattering becomes less important and the dominant effect is reflection.

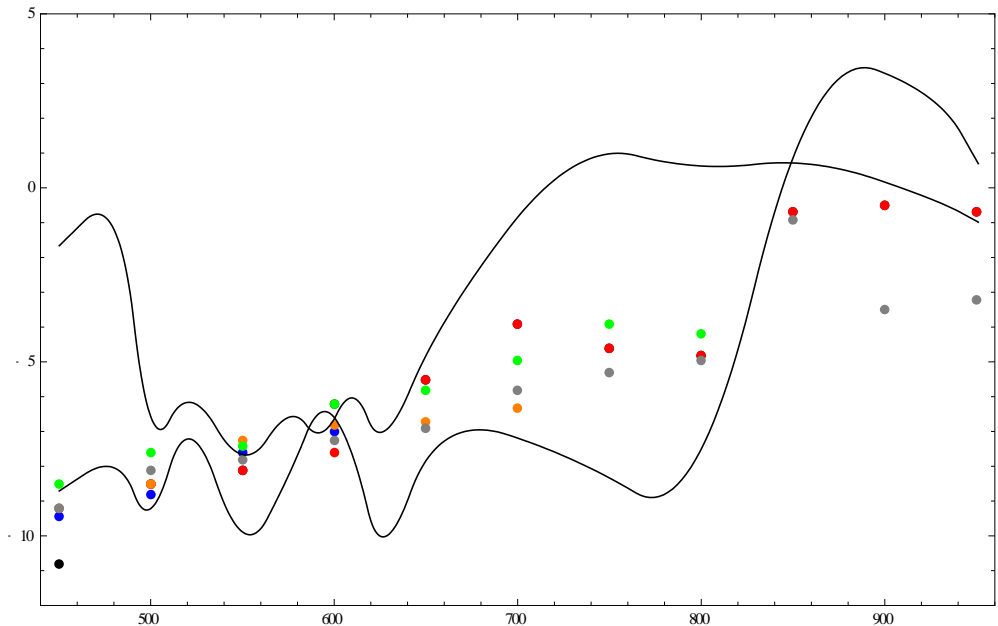

Fig. 7. Comparison between the (logarithmic scale) coefficient of absorption $K_d$ in experiments with black screen (upper curve), gray screen (lower curve) from Figs. 6 and from literature. Legend: Black dots represent data from reference [6], red dots [7], blue dots [8], orange dots [5], green dots [9], purple dots [10], and gray dots from reference [11].

**7 Conclusions**

The comparison between the two experiments with black and gray screens are presented in Fig. 7 together with the values obtained in different literature references.